\documentclass[11pt]{article}
\usepackage[dvips]{graphicx}
\usepackage{amsfonts}
\usepackage{amsmath}
\usepackage{amssymb}
\newcommand{\p}{\ensuremath{\mathrm{P}}}
\newcommand{\rp}{{\rm R}}
\newcommand{\np}{\ensuremath{\mathrm{NP}}}
\newcommand{\up}{\ensuremath{\mathrm{UP}}}
\newcommand{\npcapconp}{\ensuremath{{\np\cap\conp}}}
\newcommand{\pspace}{{\rm PSPACE}}
\newcommand{\ip}{{\rm IP}}
\newcommand{\poly}{\ensuremath{{\rm poly}}}
\newcommand{\pisnp}{\ensuremath{\p=\np}}

\newcommand{\pisnotnp}{\ensuremath{\p\neq\np}}
\newcommand{\sat}{{\rm SAT}}
  \newtheorem{theorem}{Theorem}[section]
\newcommand{\conp}{{\rm coNP}}

\title{Beautiful Structures:\\An Appreciation of the 
Contributions of Alan Selman\footnote{This appreciation
will appear, in slightly different form, 
in the Complexity Theory Column of the 
September 2014 issue of \emph{SIGACT News}.}}

\author{Lane A. Hemaspaandra\footnote{Supported in part 
by grant 
NSF-CCF-1101479.}\\
Department of Computer Science\\University of Rochester\\
Rochester, NY 14627, USA}
\date{June 17, 2014}
\begin{document}

% If a maketitle is called, the default is to not use the
% fancy page style which leads to no ACM SIGACT News footline.
% So an alternative maketitle has been provided.
% Note added 2006/9/26: but due to the switch to footers being 
% provided centrally by the head editor, via a \thispagestyle{empty}
% this is overridden right after the \SIGACTmaketitle, below.
\maketitle

% If a maketitle is called, the default is to not use the
% fancy page style which leads to no ACM SIGACT News footline.
% So an alternative maketitle has been provided.
% Note added 2006/9/26: but due to the switch to footers being 
% provided centrally by the head editor, via a \thispagestyle{empty}
% this is overridden right after the \SIGACTmaketitle, below.
%\SIGACTmaketitle
%\thispagestyle{empty}
%\vspace*{-0.7in}

%\begin{center}
%{\protect\normalsize Lane A. Hemaspaandra}\\
%{\protect\normalsize Dept.~of Computer Science, University of Rochester}\\
%{\protect\normalsize Rochester, NY 14627, USA
%}
%\end{center}

%%%% the above is for a short intro.  
%%%% and for a column with a somewhat longer intro, as will be the case
%%%% with bill's survey article probably, as i probably have my own
%%%% comments about P vs. NP, more space is needed and thus:
% This block of text is the amount of space you need to please leave for
% Lane's introduction.  This block of text is the amount of space you
% need to please leave for Lane's introduction.  This block of text is
% the amount of space you need to please leave for Lane's introduction.
% This block of text is the amount of space you need to please leave for
% Lane's introduction.  This block of text is the amount of space you
% need to please leave for Lane's introduction.  This block of text is
% the amount of space you need to please leave for Lane's introduction.

%\bigskip

% your contributions go here.

\begin{abstract}
Professor Alan Selman has been a giant in the field of computational
complexity for the past forty years.  This article 
is an appreciation, on the occasion of his retirement, of some of the most
lovely concepts and results that Alan has contributed to the field.
\end{abstract}

\section{Preface}

\begin{figure}[!h]
\begin{center}
\includegraphics[height=43mm]{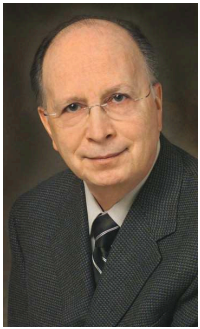}
\end{center}
\caption{Alan Selman}
\end{figure}
%\section{Introduction}
%
%\subparagraph{Alan Selman and This Issue's Column} 
As I write these words in
June 2014, it has been just over a month since the retirement celebration
for Alan Selman at the University at Buffalo's Center for Tomorrow.  
I can't think of a more fitting location for the celebration, 
given that Alan's 
technical contributions
to the field
%---in concepts, results, and proteges---
are of such beauty and insight that they are as important to the
field's future as they have been to its past and present.

Any retirement is bittersweet, but 
%as I mentioned in the previous
%column's introduction, 
Alan has mentioned that he will be keeping his hand in the field in
retirement.  That happy fact helped all of us at the celebration
focus on the sweet side
of the bittersweet event.  Warm talks and memories were shared by
everyone from the university's president, the department chair, and Alan's
faculty colleagues all the way up to the people who are dearest of all
to Alan---his postdocs and students.

The warmth was no surprise.  Alan is not just respected by but also is 
adored by those who have worked with him.  Anyone who knows Alan
knows why.  Alan is truly kind, shockingly wise, and simply by his
nature devoted to helping younger researchers better themselves and
the field.  But in fact, I think there is more to say---something far
rarer than those all too rare characteristics.  What one finds in Alan
is a true belief in---an absolute, unshakable belief in---%and
%dedication to 
the importance of understanding of the foundations of
the field.  
%That intensity of belief is far from universal even among
%theory people.

Now, one might think that Alan holds that belief as an article of
faith.  But my sense is that he holds the belief as an article of
understanding.  Like all the very, very best theoreticians, Alan has a
%great 
terrific 
% shockingly good 
intuition about what is in the tapestry of coherent
beauty that binds together the structure of computation.  He doesn't
see it all or even most of it---no one ever has.  But he 
knows it is there.  And in these days when many nontheory people 
throw experiments and heuristics at hard problems, often without 
much of a framework for understanding behaviors or evaluating
outcomes, not everyone can be said to even know
that there is an organized, beautiful whole to be seen.  Further,
Alan has such a
strong sense for what is part of the tapestry that---far more than most
people---he 
% during his career has 
has revealed the tapestry's parts and has guided his
collaborators and students in learning the art of 
%seeking to themselves reveal pieces of the tapestry.
discovering pieces of the tapestry.

And that brings us to the present article and its theme of the beauty
of the structures and the structure that Alan has revealed---the
notions, the directions, and the theorems.  For all of us whose
understanding isn't as deep as Alan's, the beauty of Alan's work has
helped us to gain understanding, and to know that that tapestry really
is out there, waiting to be increasingly discovered by the field,
square inch by square inch, in a process that if it stretches beyond
%single 
individual lifetimes nonetheless enriches the lifetimes of those involved
in the pursuit of something truly important.  To summarize Alan's
career in a sentence that is a very high although utterly deserved
compliment: Alan is a true structural complexity theorist.
% and that is perhaps the highest complement of all.

%
%Myself, I sat two of Alan's graduated Ph.D. students, Ashish Naik and
%Samik Sengupta.  Hearing how wonderfully their professional and
%personal lives are going was a real treat, but hardly a surprise.  
%And it also was no surprise

%\section{Overview}
\section{Introduction}
It would be impossible to cover in a single reasonably sized 
article all or even most 
of Alan's contributions to the field. 
So this article will celebrate Alan's career 
in a somewhat unusual way.  Given that the heart of Alan's contribution
to the field is truly beautiful structures---notions, directions and 
foundational results regarding those---this article will 
%mostly limit itself
%to 
simply present to the reader a few of those structures and point out 
(when it isn't already apparent)
%on 
%trying to 
%make clear 
why they are beautiful and important.  

We won't be trying to survey the results that are by now known about
the structures, although we'll often mention the results that Alan and
his collaborators obtained on the structures in their original work,
and we'll sometimes mention some later results.  But the core goal here is
to present the beautiful structures themselves.  

We'll do that in Section~\ref{s:structures}.  But before turning to
that, it would be a crime not to at least allude to the stunning
breadth of Alan's contributions in service and developing human
infrastructure, 
%although those important topics are not the focus of
%this article, 
and to the recognition he has received for his
scholarship and service.

Alan has trained many
%more than a dozen current and former 
Ph.D. students
(his already graduated Ph.D. advisees are 
Joachim Grollmann, John Geske, Roy Rubinstein, Ashish Naik,
Aduri Pavan, Samik Sengupta, and Liyu Zhang, and his
current Ph.D. advisees are 
Andrew Hughes, Dung Nguyen,
and 
Nathan Russell)
and postdocs (Mitsunori Ogihara,
Edith Hemaspaandra, and Christian Gla{\ss}er), 
and their contributions---both under Alan and beyond---have
been important to the field.  Those of us who were not his
students or postdocs but have had the privilege of working with Alan
have been enriched, inspired, and uplifted by his insights and
vision.  The field and his school have recognized Alan's research and
service contributions with a long list of the highest awards and most
important positions.  Alan is a Fellow of the ACM, a Fulbright
scholar, and the recipient of an Alexander von Humboldt Foundation Senior
Research Award and a Japan Society for the Promotion
of Science Invitational Fellowship.  He has been awarded the ACM
SIGACT Distinguished Service Prize, and has a long record of extensive
editorial service to the field, including being the editor-in-chief of
Theory of Computing Systems since 2001.  He is the recipient of the
State University of New York Chancellor's Award for Excellence in
Scholarship and Creative Activities, and of the University of
Buffalo's Exceptional Scholar Award for Sustained Achievement.  Alan was
acting Dean of the College of Computer Science at Northeastern
University, and chaired UB's Computer Science and Engineering 
department for six years.  
With Steve Homer, Alan wrote a computability and complexity 
textbook~\cite{hom-sel:bsecondedition:complexity}, and
he edited the Complexity
Theory Retrospectives~\cite{sel:b:retrospective,hem-sel:b:ctrII}.  
Alan was an
important part of efforts to obtain stronger government grant support
for the study of theoretical computer science.  He was instrumental in
the creation of, and served the first term as conference chair of, the
IEEE Conference on Structure in Complexity Theory (now called the IEEE
Conference on Computational Complexity), and for that won the IEEE
Computer Science Society Meritorious Service Award.

That's some record!
% to have built in just a lifetime.

\section{Beautiful Structures}\label{s:structures}
In this article, we'll present 
%just 
five of Alan's beautiful notions---concepts that Alan has introduced
or very substantially advanced.  That is such a small number relative
to the dozens of topics that Alan has contributed to that we'll be
skipping topics that for almost any other person would in and of themselves
be the highlight of an entire career.  In fact, we won't even be trying to
pick off the ``top five'' notions, but rather will just be selecting a
set of five very lovely notions.

For example, we'll skip right over the seismic contribution of Ladner,
Lynch, and Selman~\cite{lad-lyn-sel:j:com} to the definitions of and
understanding of the relative powers of the rich range of
polynomial-time reducibilities that are so widely used today.  

And
similarly, since Alan and his collaborators will soon be surveying issues
related to these topics in a \emph{SIGACT News} Complexity Theory
Column~\cite{gla-hug-sel-wis:jinprep:disjoint}, 
we also will leave untouched all issues related to disjoint
pairs, promise problems, and propositional proof systems, even though
Alan's importance there is great, stretching across more than
a half dozen papers, 
from the seminal 1980s work
of Even, Selman, and Yacobi~\cite{eve-sel-yac:j:promise-problems}
to the remarkable 2010s work of 
Hughes, Pavan, Russell, and Selman~\cite{hug-pav-rus-sel:c:promise-problems}.  

Alan and his collaborators 
have in recent decades resolved a
long-open issue, bringing unity to the understanding of mitoticity,
completeness, and autoreducibility for many central complexity
classes.   Among their advances is that 
every (many-one
polynomial-time) NP-complete set is (many-one polynomial-time)
autoreducible, and (many-one polynomial-time) autoreducibility and
(many-one polynomial-time) mitoticity coincide.  So, for example,
every NP-complete $A$ set is so repetitively structured that there
exists a $\p$ set $B$ such that $A \,\cap \,B$ and $A \,\cap\, \overline{B}$ are
infinite and (many-one polynomial-time) equivalent, and so certainly
are themselves each (many-one polynomial-time) NP-complete.  Also,
Alan and his collaborators have brought great light to the extent
to which this type of behavior persists or fails for other types of
reductions, such as for reductions more flexible than many-one
reductions, or reductions in the logspace world.  And yes, you've
guessed it, we also won't be covering any of that work here, since
happily Alan and his collaborators half a decade ago wrote for the
\emph{SIGACT News} Complexity Theory Column
a survey article on the work in this line up to that
point~\cite{gla-ogi-pav-sel-zha:j:mitoticity-autoreducibility-survey},
although if you are interested (as well you should be!),~please don't
miss their recent work on this line in ICALP-13 and
STACS-14~\cite{gla-ngu-rei-sel-wit:c:autoreducibility-logspace,ngu-sel:c:nonautoreducibility-nexp}.

$\np \cap \conp$---not just itself, but as it functions when accessed
through relativization (e.g., $\p^\npcapconp = \npcapconp$ and 
$\np^\npcapconp = \np$), and in its cousin known as ``strong''
computing, and in the so-called strong nondeterministic
reductions~\cite{sel:j:enumeration-reducibility,lon:j:nondeterministic-reducibilities}
based on that cousin---plays an important role in complexity theory.
Alan was a central player in 
the key early work that built this collection of 
concepts~\cite{sel:j:structure-np,sel:j:enumeration-reducibility,%
% this 1979 paper is related to lowness of NP cap coNP:
sel:j:pselective-tally,%
bra:j:cryptocomplexity,%
lon:j:nondeterministic-reducibilities,sch:j:low}.
%much of it was done by
%Alan and his collaborators.  
And we won't cover that here.  

We also won't cover what is now called the left-set technique,
which was created by Alan~\cite{sel:j:natural} and which for example 
was used 
to devastating effect by Ogihara and Watanabe~\cite{ogi-wat:j:pbt}
in their work showing that if any NP-complete set polynomial-time 
bounded-truth-table reduces to a sparse set, then $\pisnp$.

There are many other beautiful themes, notions, and results in
Alan's work, which we not only won't cover below but which we also
haven't mentioned above.  In fact, it should already 
be clear what we won't cover
about Alan's work is enough to fill three or four extremely
satisfying, productive, important careers.  But were we to go on
listing the important and lovely notions due to Alan that we \emph{won't}
cover, there would be no room left to actually cover any structures that Alan
developed.  So let us move right on to our select five.

As a reminder, for these notions we won't at all be doing a survey of
what is known, but rather we will be trying to convey what the notion is,
why it is beautiful, and why its introduction by Alan and his
collaborators was important.  And a shorthand, this article will sometimes
say ``Alan and his collaborators'' or even perhaps ``Alan''
when speaking of 
work
%a paper 
by Alan joint with others; 
this is not in any way meant as a slight to those
collaborators, but is simply since the focus of this article
is on Alan.  The citation labels
(e.g.,~``[BLS84]'') will generally make it clear to the reader when
we are employing this shorthand.

\subsection{P-Selectivity: Choosing Which Path to Take When the 
Road Forks}
Search issues are important not just in the area of Theory.  AI
researchers also are intensely focused on how to explore spaces.

So suppose you come to a fork in the road.  There might be a treasure
down one or both of the paths---or neither might hold a treasure.
Which way should you go?

One of Alan's beautiful structures addresses the issue of when
polynomial-time functions can help you solve the above problem.  After
all, you don't really need to know whether a given road has a
treasure.  That would be great to know, but maybe it is too much to
hope for or too computationally expensive.  Happily, all you really need to be
able to do in the above situation is to choose one of the roads such
that it is true that \emph{if either of the roads holds a treasure, then
  the road you choose holds a treasure}.

In a quite amazing series of
papers~\cite{sel:j:pselective-tally,sel:j:some-observations-psel,sel:j:ana,sel:j:reductions-pselective},
Alan introduced and broadly explored the notion of P-selectivity,
which captures precisely the above issue.  We all know what polynomial
time~($\p$) is.  A set $A$ is in P exactly
if there is a polynomial-time machine that
accepts on each string in $A$ and rejects on each string in
$\overline{A}$.  That is, one has a polynomial-time decision algorithm for
membership in the set.  For P-selectivity, one is required to have a
polynomial-time semi-decision algorithm for the set.  
In particular---and here we use 
a particular one of the various equivalent definitions of this quite robust
concept---a set $A$ is said to be P-selective exactly if there is a
polynomial-time function $f$ that takes as its input two arguments,
$x$ and $y$, and has the property that, for each $x$ and $y$, 
$$ f(x,y) \in \{x,y\} \land (\{x,y\}\cap A \neq \emptyset \implies 
f(x,y)\in A).$$

What this says is that a set is P-selective exactly if there is a
polynomial-time function that, given any two elements, always chooses
one of them, and if at least one of them is in the set, the one it
chooses is in the set.  One sometimes hears this described by saying
that the
function chooses the element ``more likely''---or, better, ``no less
likely''---to be in the set.  This is a fine 
description, at least in the ``no less
likely'' version, as long as one keeps in mind that the probabilities
here are all zero and one.

P-selectivity is 
%in essence 
capturing the notion of wise search---being
able to decide which way to go at forks in the road.  It also is one
of a great variety of notions (such as
Almost Polynomial Time~\cite{mey-pat:t:int}, 
P/poly~\cite{kar-lip:c:nonuniform}, 
P-closeness~\cite{sch:j:closeness}, 
Near Testability~\cite{gol-hem-jos-you:j:nt}, Nearly Near
Testability~\cite{hem-hoe:j:nnt}, etc.)\ that try to capture a wider
range of sets than P does, yet to still have some natural
polynomial-time action at their core.

Although this lovely notion, P-selectivity,
was introduced by Alan, Alan's seminal papers are
quite open as to his inspiration.  The P-selective sets (which are
also sometimes called the semi-feasible sets, since they are the sets
having polynomial-time semi-decision algorithms) were inspired by
Jockusch's notion from recursive function theory of the semi-recursive
sets~\cite{joc:j:semi}.  In his career, Alan has often drawn on his
broad understanding of recursive function theory to improve 
computer science's pursuit of 
complexity-theoretic insights.  This is of 
great value, given that so
much of complexity theory---from the polynomial 
hierarchy~\cite{mey-sto:c:reg-exp-needs-exp-space,sto:j:poly} to
reductions to completeness to 
the Isomorphism Conjecture 
to immunity and bi-immunity to oracles 
and
much more---%has recursion-theoretic roots.
% cursive function theory.
%can be viewed as inspired by recursive function theory.
%is built in the image of recursive function theory.
%is sharped by th
%has recursion-theoretic analogues.
is inspired by recursive function theory.

The present article focuses on presenting the beautiful structures
themselves, rather than surveying everything known about them.  But in
this case it is important to note that although the P-selective sets
have been extensively studied since Alan's original series of papers
(and indeed, there is even a monograph completely devoted to 
selectivity theory~\cite{hem-tor:b:semifeasible-computation}), Alan's
original series of papers already went 
%almost unbelievably 
breathtakingly far in exploring this concept.  

For example, Alan's original papers already proved that there are
P-selective sets that are not in P\@.  Indeed, he (see also
Ko~\cite{ko:j:self-reducibility-CAREFUL-selman-did-left-cuts-first-see-comment}
regarding a more flexible type of left cut) showed that the left cut
set 
(don't worry if you don't know what that term means) of any real
number is P-selective.  It follows from this
%in some sense 
%shows 
that there are 
% in a certain sense
arbitrarily hard P-selective sets, e.g., there are P-selective sets
that are so wildly undecidable that they aren't even in the
arithmetical hierarchy.  

Alan also showed that if even one NP-hard set is P-selective, then 
%P equals $\np$; the proof is a lovely application of the
$\pisnp$.  
The proof is a lovely application of the
self-reducibility of satisfiability.  
Going back to the parallel with
wise search, this basically says that for NP-like search spaces that
can be naturally cut in half repeatedly, having a P-time wise search
algorithm is just too much to hope for.  The proof is so crisp that it
is worth sketching here.  Suppose that some NP-hard set $A$ is
P-selective.  Here is a polynomial-time algorithm for satisfiability.
Given a boolean formula whose satisfiability we want to test, set
its lexicographically
first variable to true, get the resulting formula,
and by $A$'s NP-hardness transform that into a question about $A$.  Do
the same for our original formula with its first variable set to
false.  Take the outputs of these two processes, and use them
as the inputs to the hypothesized 
polynomial-time P-selector function for $A$.  That function
will output one of them, and based on that choice, we'll know either
``If the formula is satisfiable then there is a satisfying assignment
in which the first variable is set to true,'' or ``If the formula
is satisfiable then there is a satisfying assignment in which the
first variable is set to false.''  So we take whichever
assignment was just suggested to us by the selector, and we stick with
it and similarly assign the second variable, again using
the selector to fix it as being true or false, and so on until all
the variables are assigned.  At the end, we get to a single assignment
such that if there is any assignment that satisfies the original
formula, then \emph{that} assignment satisfies the original formula.  So we
see if that one assignment satisfies the formula.  If it does the
formula is satisfiable, and otherwise the formula
is not satisfiable.  Lovely.

The results mentioned above are just a few examples of the broad
investigation Alan carried out.

The facts that P-selective sets can by arbitrarily hard, and
can't be NP-hard unless 
$\pisnp$, might lead one to say that this
class is terribly far beyond P\@.
However,
using a tournament-inspired divide-and-conquer framework,
Ko~\cite{ko:j:self-reducibility-CAREFUL-selman-did-left-cuts-first-see-comment}
% promptly after Alan's original work, 
proved that each P-selective set is
information-wise very close to P\@. In particular, for each
P-selective set there is
a polynomial-time algorithm that given a polynomial amount of extra
information based only on the 
\emph{length} of the input string can correctly accept
the set.  In the jargon, each P-selective set is in the complexity
class P/poly, or equivalently, each P-selective set has small circuits.

Thus the P-selective sets have two faces: They can be so hard as to be
undecidable, 
%and yet information-wise they float within a hair's
%breadth of being in P\@.  
yet only a hair's breadth of information keeps them from being in $\p$.

Beautiful structures often attract much attention, and that has
certainly been the case with the P-selective sets.  Much of that
attention has been devoted to generalizing and extending 
concepts and results from 
Alan's seminal papers.  For example, although Alan showed that no
NP-hard set (i.e., no set NP-hard with respect to many-one
polynomial-time reductions) can be P-selective unless 
%P equals NP,
$\pisnp$,
there followed an intense and productive research line to see whether that
claim could be extended beyond many-one reductions to more flexible
types of reductions such as bounded-truth-table and
truth-table reductions, see
\cite{bei-kum-ste:j:approximable-sets,ogi:j:comparable,agr-arv:j:psel},
and whether Alan's just-stated result analogously
extends to other complexity classes, and what
holds for nondeterministic variants of selectivity.  See
\cite{hem-tor:b:semifeasible-computation} for a survey of work along
those lines and more generally for a survey of the broad research
stream---including important later work 
% e.g., Naik-Selman
by Alan---launched by Alan's beautiful structure known as
P-selectivity.  As a final comment, in Section~\ref{s:hnos} of this
article we'll soon see how selectivity theory
%, in a totally
%unexpected way, 
surprisingly resolved an important, yet seemingly unrelated,
question about removing the ambiguity of
nondeterministic functions.

\subsection{Much Ado about Functions: Decision Problems Are Not the Only Game in Town}

As everyone knows, theoretical computer science is largely built
around the complexity of decision problems.  After all, NP is (at
least if one stays away from certain textbooks!)~a class of decision
problems.  How could there be anything wrong with this approach?
After all, satisfiability for example clearly has the property that
its search and its decision versions are polynomial-time Turing
interreducible.

Alan is one of the people who from the very beginning recognized the
tremendous importance of studying functions directly.  There 
are many compelling reasons to study functions directly.  
Historically, Alan was probably
primarily motivated by the extremely interesting role and richness 
of nondeterministic function classes---and his work there was seminal---and
perhaps also by his expertise in one-way functions.

Additionally, experts such as Alan have always known that even for
deterministic classes, the justification given above for focusing on
decision problems has weaknesses.  Satisfiability as used above is not
as canonical as it often is.  Indeed, it has since been shown that
unless 
%P equals NP 
$\pisnp$ some NP-complete sets are
not 
self-reducible~\cite{fal-ogi:c:self}, which is relevant
since self-reducibility is central in reducing search problems to 
their natural associated decision problems.\footnote{We say ``natural''
since for any NP-like search problem one can build some decision 
problem to which it Turing-reduces.  But that is not the focus here.}
But more crucially, Turing
reductions are quite powerful, and so may give a blurred view of the
actual complexity of the objects involved.

Early on, Alan and his collaborators Book and Long
(\cite{boo-lon-sel:j:quant}, see also \cite{adl-man:c:red,boo-lon-sel:j:qual})
introduced
a rich range of polynomial-time nondeterministic reductions and
studied their properties and relationships.  Alan
%, as he has
%often done in his career, made special efforts to 
%make broadly 
%clear
%and available 
also accessibly spread the word about the
importance of taking a function-based view,
%.  He did so for example in
through his papers on ``A Taxonomy of Complexity Classes of
Functions''~\cite{sel:j:taxonomy} and ``Much Ado about
Functions''~\cite{sel:c:much-ado} (see also his 
survey paper on one-way functions~\cite{sel:j:one-way}).  

As 
he was writing those papers, Alan was also 
obtaining new insights into search (a type of function 
version of problems) versus decision, e.g., his paper
``P-Selective Sets and Reducing Search to Decision vs.\
Self-Reducibility''~\cite{hem-nai-ogi-sel:j:pselective} (and yes,
there is P-selectivity playing a role again!),~and he 
also interestingly studied the role of functions as outputs 
of oracles~\cite{fen-hom-ogi-sel:j:using-oracles}.

These days, function versions of problems play such a large role that
it is easy to forget that thirty years ago it wasn't yet clear that
that would ever be the case.  Alan's pioneering stress on functions,
quirky for its time, was quite prescient.  

\subsection{Selectivity and Functions Team Up: Is Finding All
Solutions Easier than Finding One Solution?}\label{s:hnos}

Is finding all solutions to a problem easier than finding one
solution?  One is tempted to answer: Never!  After all, if one can
find all solutions, then one can simply take (for example) the
lexicographically smallest solution, and one has found one solution
(if one exists).

Surprisingly, the reasoning just used falls apart in a
nondeterministic context.  Let us see this.  And then let us see what
Alan did about this through bringing together the theory of functions
with a generalization of the theory of P-selectivity.

Consider a nondeterministic polynomial-time Turing machine that on
each path either rejects or accepts.  We will consider the machine to
be computing a partial, multivalued function (which in mathematics one would
probably call a relation, but in complexity theory the term function is
often used even for such multivalued objects).  Namely, each
rejecting path isn't considered to contribute anything to the
function.  But whatever string is on the first work tape of the
machine on nondeterministic paths that accept is considered to be an
output of the function on the given input.  This notion is 
what is called an 
NPMV function (a nondeterministic polynomial-time multivalued 
function).  The class NPMV was introduced by Book, Long, and
Selman~\cite{boo-lon-sel:j:quant} in their seminal work on 
nondeterministic functions.

Let us make the nature of NPMV clear by giving a very important
example.  Consider the nondeterministic polynomial-time Turing
machine~(NPTM) that on input $f$, interprets $f$ as a boolean formula,
nondeterministically guesses an assignment for the variables of $f$
and writes that assignment on its first work tape, and then
deterministically uses its other tapes to check whether the assignment
satisfies $f$, and if it does the machine (on that nondeterministic
computation path) accepts and if not the machine (on that
nondeterministic computation path) rejects.  Notice that this machine
on input $f$ outputs all satisfying assignments of $f$.  That set
could be exponentially large, but that isn't a problem here---the
output set is in effect distributed among all the paths , and so
it isn't ever stuffed into a single, giant output string.  This function
is called $f_{\sat}$.

An NPSV function (a nondeterministic polynomial-time single-valued
function) is exactly the same notion as an NPMV function, except 
NPSV functions must in
addition satisfy the property that on each input the cardinality of
the output set is at most one.  So if no path accepts, that is fine, as
the cardinality of the output set is zero.  And if one or more paths
accept, that is also fine, as long as every one of those accepting
paths has precisely the same string on its first work tape, since then
that string will be the one and only output.

A central concept in the study of multivalued functions is the notion
of a refinement.  $f_2$ refines $f_1$ exactly if on each input $x$, 
\begin{enumerate}\item $f_2$
outputs at least one value if and only if $f_1$ outputs at least one
value, and 
\item every output of $f_2$ is an output of $f_1$.  
\end{enumerate}

We saw above that $f_\sat$ is an NPMV function.  But does it have an
NPSV refinement?  Put another way:
As mentioned 
above, it is easy for NP function-computing
machines to find all solutions of an input boolean formula.  But can an NP
function-computing machine find \emph{one} solution of an input
boolean formula
(when one exists)?  
% Taking $f_\sat$ again as our example, and trying
%to refine it to an NPSV function, 
This question is asking whether there is an
NPTM that for unsatisfiable formulas has all its paths reject, and for
each satisfiable formula has at least one path that computes a
satisfying assignment (and accepts) and every accepting path must
compute the same satisfying assignment as every other accepting path.

It will now be clear why the trick that works in the deterministic case
seems unhelpful here.  In the deterministic case
we took all the solutions and output the
lexicographically smallest.  But for an NPMV function to do that, at
least in the most obvious way, a path would have to be able to figure
out whether the value it would like to output is such that every other
path that would like to output a value would in fact like to output a
lexicographically equal-or-larger value (and in that case our path 
would go ahead and output its value, and otherwise would kill itself 
off by rejecting).  But figuring \emph{that} out seems to take an extra
quantification that our machine doesn't have in its arsenal.  
Big picture, NPSV is in some sense 
asking for a strong degree of coordination
among paths that have no way of communicating with each other.

Of course, the previous paragraph is just an intuitive handwave, 
not a
proof.  Proving that NPMV functions don't all have NPSV refinements
(unless the polynomial hierarchy collapses) required a 
surprising twist.  Alan and his collaborators had developed
nondeterministic analogues of
P-selectivity~\cite{hem-hoe-nai-ogi-sel-thi-wan:j:np-selective,hem-nai-ogi-sel:j:refinements}.
Although at first one might think that selectivity
has little to do with NPMV and
NPSV functions, nondeterministic selectivity and its connection with
nonuniform classes such as $(\np\cap\conp)/\poly$ turned out to be
exactly the tool Alan and his collaborators needed to prove that if
$f_\sat$ has an NPSV refinement (equivalently, if every NPMV function
has an NPSV refinement), then the polynomial hierarchy collapses to
$\np^\np$~\cite{hem-nai-ogi-sel:j:refinements}.  

% That same collapse holds 
A collapse of the polynomial hierarchy to $\np^\np$ still can 
be obtained
even if one merely assumes that 
one can refine at-most-2-valued NPMV functions to NPSV
functions~\cite{hem-nai-ogi-sel:j:refinements}.  Later work about
nonuniform classes let one conclude from Alan's work a slightly more
extensive collapse in both these
cases~\cite{cai-cha-hem-ogi:j:s2-and-lowness}, in particular
collapsing the polynomial hierarchy to 
$\rm S_2^{\np\cap\conp}$.  Alan
and his collaborators 
%later 
also explored the 
%still-open 
question of, for
$k \geq 2$, what collapses would occur if one could reduce
at-most-$(k+1)$-valued NPMV functions to at-most-$k$-valued NPMV
functions, and
obtained polynomial-hierarchy collapses to $\np^\np$ for each such
case~\cite{nai-rog-roy-sel:j:npkv}.  
%As an aside, 
Strengthening those
collapses to a collapse to $\rm S_2^{\np\cap\conp}$ remains open to
this day and is a quite interesting challenge that has stymied many a
graduate student.  See also~\cite{hem-ogi-wec:j:npmv-refinement-and-lowness}
for some cases where refinements in fact are possible.

In summary, Alan didn't define nondeterministic functions and 
selectivity theory for the purpose of shedding
light on whether one could reduce many solutions to one solution.
However, his notions were so beautiful and flexible that they were
central in the resolution of that issue.

\subsection{Positive Relativization: Getting Real-World Consequences from
  Oracles}

Informally put, 
%Recall that 
relativizing by an oracle $A$ means
giving all machines unit-cost access to $A$, i.e., machines can as often as
they like write a string onto a query tape and immediately the 
machine will be
told whether that string is a member of $A$.  In some sense, relativization
changes the universe's ``ground rules'' about what information is
available to computing machines, but does so fairly---all machines now
have access to $A$.

Experienced complexity theorists may be a bit surprised by what we've
chosen as our example of a beautiful contribution by Alan to 
the theory of relativization.  

After all, one of Alan's earliest works is the famous 1979 paper ``A
Second Step Toward the Polynomial Hierarchy'' with
Baker~\cite{bak-sel:j:step2}, which showed that there is an oracle
relative to which the second and third levels of the polynomial
hierarchy separate.  The proof is extremely clever and
powerful---employing what one might dub a nested double contradiction
architecture.  How amazing that proof was is made clear by how long it
took to take the ``third step''; that didn't occur until the work many
years later of Yao~\cite{yao:c:separating} and
H{\aa}stad~\cite{has:thesis:small-depth}, which drew on different
techniques and separated the entire polynomial hierarchy.

Structural complexity theorists will remember well the important
paper by Homer and Selman~\cite{hom-sel:j:iso} that built a
relativized world in which all $\np^\np$-complete sets were
polynomial-time isomorphic.  Not too many years later, Fenner,
Fortnow, and Kurtz~\cite{fen-for-kur:jisomorphism-holds},
surely encouraged by that paper, obtained the
same result for NP itself, thus directly speaking to the relativized
version of 
Berman and Hartmanis's Isomorphism 
Conjecture~\cite{ber-har:j:iso}.  Rogers~\cite{rog:j:iso-one-way}
even put a great cherry on top of that, by obtaining isomorphism while 
not killing off one-way functions---and we'll speak more 
about those in Section~\ref{s:one}, since 
Alan's work on them is seminal.

But as beautiful and important as those results of Alan's are, at
least as beautiful,
if perhaps less well known these days, 
is the
pioneering work by Alan and his collaborators on positive
relativization.

What is the biggest perceived weakness of relativization theory?  It
probably is that it often seems that one can do almost anything in
relativized worlds, and that doing so has little connection with the
real world.  For example, Baker, Gill, and
Solovay~\cite{bak-gil-sol:j:rel} famously showed that there are
oracles making P equal to NP, and that there are oracles making P not equal
to $\np$.  Does this resolve the P versus NP problem in the real
world?  Fat chance.  

Those oracles actually do tell us something quite
important, namely, they tell us that no proof technique that
relativizes can prove either $\p=\np$ or $\pisnotnp$.  And so there has
been quite a bit of research regarding proof techniques such as
arithmetization (see~\cite{bab-for:j:arithmetization}) and results
such as $\ip = \pspace$~\cite{sha:j:ip} that seem not to 
relativize~\cite{for-sip:c:ip-conp}.  
An interesting perspective on this is given by
Hartmanis et
al.~\cite{cha-cha-har-ran-roh:j:revisionist}, who argue that there
have been nonrelativizing proof techniques for a long time; see
also~\cite{cha-cho-god-har-has-ran-roh:j:random-oracle}.  

In any case,
%Nonetheless,
the fact that relativization results such as 
those of Baker, Gill, and
Solovay~\cite{bak-gil-sol:j:rel} fail to resolve the analogous 
real-world problems is often presented as a clear weakness of 
studying oracles.

Positive relativization provides an utterly lovely response to this
weakness: \emph{We should find cases where obtaining an 
oracle result would imply
real-world results.}  Positive relativization 
as such
% the as such is because Book in Juris's book cites his 1981 bounded
% query paper as the start of it.
was introduced and explored 
by Alan, Ron Book, and their collaborators~(\cite{boo-mei-sel:j:pos}, 
see also~\cite{boo:j:bounded-query,boo-lon-sel:j:quant,bal-boo-lon-sch-sel:c:sparse-oracles,boo-lon-sel:j:qual,lon-sel:j:sparse,boo:c:tp,boo:b:restricted}).

Let us illustrate positive relativization by one of its most
striking examples.  
%(Warning: The terminology people use differs.
%For example, Book~\cite{boo:b:restricted} 
%in part uses the terms ``negative
%relativization'' and ``restricted relativization'' for what we here
%will simply group together 
%as being positive relativization.)  
(Note: 
%The terminology people use differs.
%For example, 
What we here speak of as ``positive relativization''
also sweeps in what 
Book~\cite{boo:b:restricted} 
distinguishes as ``negative relativization.'')
%and ``restricted relativization'' for
% what we here
%will simply group together 
%as being positive relativization.  
%relativization.''
%
% (Warning: The terminology people use differs.
% For example, Book~\cite{boo:b:restricted} 
% in part uses the terms ``negative
% relativization'' and ``restricted relativization'' for what we here
% will simply group together 
% as being positive relativization.)  
Suppose we claim that
there is sparse oracle relative to which the polynomial hierarchy
collapses.  Are you thrilled?  Perhaps not, since that might not say
anything about the real world.  Suppose we claim that there is sparse
oracle relative to which the polynomial hierarchy is infinite.  Now are
you thrilled?  Perhaps again not, since that might not say anything
about the real world.  You're seeming pretty hard to thrill, my friend.

But wait.  These things \emph{would} say something about the real
world.  That is because Balc{\'{a}}zar, Book, Long, Sch{\"{o}}ning,
and
Selman~\cite{bal-boo-lon-sch-sel:c:sparse-oracles,bal-boo-sch:j:sparse,lon-sel:j:sparse} proved the following result.

\begin{theorem}
\begin{enumerate}
\item The polynomial hierarchy collapses if and only if there is 
a sparse oracle relative to which the polynomial hierarchy 
collapses.
\item The polynomial hierarchy is infinite if and only if there is 
a sparse oracle relative to which the polynomial hierarchy 
is infinite.
\end{enumerate}
\end{theorem}

This theorem tells us 
we should care, and indeed be thrilled, if someone
%claims to have
has a sparse oracle that makes the polynomial hierarchy
infinite, or has a sparse oracle that makes the polynomial
hierarchy collapse.  Indeed, that 
%person 
%is either a crackpot or
person---thanks to the groundwork of Alan just presented---will have
so changed the real-world 
landscape of complexity that he or she is sure to win a
Turing Award.

So positive relativization links oracle results to real-world
collapses and separations.  And that is a lovely thing to do.  By now,
a broad range of positive relativizations are known, involving
issues ranging from sparse sets (as above) to tally sets
(historically the earliest form of what now is called positive
relativization), to number of queries, to aspects of the form and
structure of the querying.  A cynic might say that such results just
tell us which oracle results are too hard to hope to ever get, since
they would resolve major real-world issues.  But an optimist might say
that these offer an extra potential path to resolve those major
real-world issues.  

The path isn't a universal one.  
For example, Alan's
student Roy Rubinstein writing ``On the Limitations of Positive
Relativization''~\cite{hem-rub:j:positive}
showed that for ``semantic'' classes such as UP, 
$\rp$, and $\np
\cap \conp$, positive-relativization attempts with
tally oracles fail even 
though the analogous 
positive-relativization results are well known to succeed for NP and the
polynomial hierarchy's levels.

But even if
not universal, the path of positive relativization is certainly a
beautiful insight, especially since it has been broadly applied to 
the centrally important class~$\np$.

\subsection{One-Way Functions: A Complexity-Theoretic Characterization of
  Their Existence}\label{s:one}

For the final of our five beautiful structures 
sculpted by Alan, we take the rigorous
definition of the notion of a (complexity-theoretic) one-way
function, and the complete characterization 
(in terms of the collapse and noncollapse of important
complexity classes)
of whether one-way
functions exist.  Alan did this work with his first Ph.D. 
student, Joachim Grollmann~\cite{gro-sel:j:complexity-measures},
and this was also achieved independently 
by 
Ko~\cite{ko:j:operators} (see also~\cite{ber:thesis:iso}).

Many readers have probably already seen the definition and theorem
alluded to above, since they can be found in such excellent complexity
texts as for example Papadimitriou~\cite{pap:b:complexity} and Du and
Ko~\cite{du-ko:b:complexity}.  But since it is well worth everyone
knowing, let us quickly
%, informally 
give the definition and the
theorem.\footnote{These frame and explore the notion of
  complexity-theoretic one-way functions.  More demanding definitions
  of one-way functions---focusing on the success probability of
  probabilistic inverters---are important in cryptography.
  It would be wonderful to prove that such 
  extremely strong one-way functions exist.  However, 
  even the easier step of 
  proving that complexity-theoretic one-way functions 
  exist would---by the characterization result soon
  to be stated and the fact that 
  $\up\subseteq\np$---separate P from $\np$ and win the 
  prover a million-dollar Clay Mathematics Institute 
  prize~\cite{cla:url-hacked:p-vs-np-prize}.
  So achieving the ``easier'' step is quite important although 
  surely not easy.}

A (total) function $f$ is a (complexity-theoretic) one-way function
if and only if 
\begin{enumerate}
\item $f$ is polynomial-time computable,
\item $f$ is polynomially honest (i.e., there is a polynomial $q$ such
  that for each $x$, $q(|f(x)|) \geq |x|$; informally put, $f$ doesn't
  shrink its inputs more than polynomially much),
\item $f$ is an injective (i.e., one-to-one) function,
and 
\item $f$ is not polynomial-time invertible (since we're 
concerned here only with total, injective functions, we may 
use as our definition of this that 
% for non-1-to-1 this is too simplistic.
for any
  polynomial-time function $g$, there exists an $x$ such that $g(f(x))
  \neq x$).
\end{enumerate}

Valiant's~\cite{val:j:checking} class UP is the class of all 
NP sets that are accepted by some NPTM that on no input has more than
one accepting path.  

The characterization theorem that brings together complexity
classes and the existence of one-way functions is the following:
One-way functions exist if and only if $\rm P \neq
UP$~\cite{ko:j:operators,gro-sel:j:complexity-measures}.  The proof of
this theorem elegantly goes back and forth between the world of
one-way functions and the world of nondeterministic Turing machines.

This concept and characterization naturally inspired much related
work.  (See~\cite[Chapter~2: ``The One-Way Function
Technique'']{hem-ogi:b:companion} for a survey-like treatment,
including proofs, of the characterization mentioned above 
and the related results mentioned in this paragraph.)  For example,
researchers have looked at (what in effect is the study of)
$k$-bounded-ambiguity one-way functions (which interestingly enough
stand or fall together with 
%vanilla 
Alan's notion of one-way functions, thanks to a
nice induction proof of Watanabe~\cite{wat:j:hardness-one-way}), and
for multi-argument one-way functions one can study algebraic
properties such as associativity and commutativity (but again such
functions turn out to stand or fall together with 
%vanilla 
Alan's notion of one-way
functions~\cite{rab-she:j:aowf,hem-rot:j:aowfs}).  All such work
is clearly indebted to the beautiful, seminal work of Alan, his
student Grollmann, and Ko.

\section{Conclusion}
Anyone who has read Alan's articles knows that Alan always knows the
exact right number of words to use to motivate, explain, and develop a
concept.  So to conclude, let us try to take a page (or a quarter of a page) 
from Alan and keep
things as focused as possible, so that the points this article has been
trying to make speak clearly: Alan's work stands out as extraordinary
in that it has introduced and powerfully explored a tremendous number
of utterly beautiful structures.  We spoke at the start of the
article about the elusive tapestry of coherent beauty that binds
together the structure of computation.  Alan's career has been 
very successfully devoted to revealing parts of that tapestry.  Beyond
that, Alan has been boundlessly generous and inspirational to his many
younger collaborators and has been a leader in service to the field.
So with warmest thanks for so very much, and comforted by the
knowledge that Alan intends to keep his hand in the field, let us wish
Alan the most wonderful of retirements.

\bibliographystyle{alpha}  
%\small
%\bibliography{gry}  
\newcommand{\etalchar}[1]{$^{#1}$}

\end{document}